\documentclass[runningheads]{llncs}
\usepackage[T1]{fontenc}
\usepackage{booktabs} 
\usepackage{graphicx}
\usepackage[usenames]{color}
\usepackage{amsmath}
\usepackage{comment}
\usepackage{bbm}
\usepackage{amssymb}

\usepackage{url}
\date{\vspace{-5ex}}
\usepackage{subcaption}
\usepackage{pdfpages}
\usepackage{enumitem}
\usepackage{parskip}
\usepackage{array}
\newcolumntype{L}{>{$}l<{$}}
\newcolumntype{C}{>{$}c<{$}}
\newcolumntype{R}{>{$}r<{$}}

\title{Conv-NILM-Net, a causal and multi-appliance model for energy source separation\thanks{Supported by Accenta.ai}}

\begin{document}

\author{Simo Alami C.\inst{1,2}\orcidID{0000-0002-9091-7095 } \and
Jérémie Decock\inst{2} \and Rim kaddah \inst{3} \and Jesse Read\inst{1}\orcidID{0000-0002-1013-6724 }}
\authorrunning{S. Alami C. et al.}
%
\institute{Ecole Polytechnique, Palaiseau, France \and
Accenta, Paris, France \and IRT SystemX, Palaiseau, France}
%
\maketitle              

\begin{abstract}

Non-Intrusive Load Monitoring (NILM) seeks to save energy by estimating individual appliance power usage from a single aggregate measurement. Deep neural networks have become increasingly popular in attempting to solve NILM problems. However most used models are used for Load Identification rather than online Source Separation. Among source separation models, most use a single-task learning approach in which a neural network is trained exclusively for each appliance. This strategy is computationally
expensive and ignores the fact that multiple appliances
can be active simultaneously and dependencies between them. The rest of models are not causal, which is important for real-time application. Inspired by Convtas-Net, a model for speech separation, we propose Conv-NILM-net, a fully convolutional framework for end-to-end NILM. Conv-NILM-net is a causal model for multi appliance source separation. Our model is tested on two real datasets REDD and UK-DALE and clearly outperforms the state of the art while keeping a significantly smaller size than the competing models. 

\end{abstract}

\section{Introduction}


In 2018, 26.1\% of the total energy consumption in EU was attributed to households. This consumption mainly serves a heating purpose (78.4\%). Moreover, most of the residential energy consumption is covered by natural gas (32.1\%) and electricity (24.7\%), while renewables account for just 19.5\% \cite{stat}. However, as solar and wind generation rely on weather conditions, challenges due to intermittent generation have to be solved, and solutions for energy management such as demand response and photovoltaic (PV) battery management can play a key role in this regard. Machine Learning has proven to be a viable solution for smart home energy management \cite{RL_survey}. These methods autonomously control heating and domestic hot water systems, which are the most relevant loads in a dwelling, helping consumers to reduce energy consumption and also improving their comfort. 


An efficient energy management system has to take into account users habits in order to anticipate their behaviour. However, comfort is hard to quantify as it remains purely subjective. We argue that in an energy management context, the users are the only ones that can offer a proper evaluation of their own comfort. Hence, a satisfactory hypothesis is to consider that their current behaviour and habits are the ones that optimise their comfort. Therefore, an efficient energy management system is one that can anticipate users habits, optimise consumption levels (for example by deciding which source to use, temperature settings etc.) while offering solutions that alter users known habits as little as possible. 


Learning users' habits in a household is a hard problem mainly regarding data acquisition. The possible behaviours are diverse, if not unique, while monitoring inhabitants is not acceptable as it is a privacy infringement. From an energy provider perspective, the only available information is the household's total power consumption. A solution is therefore to decompose this consumption into the consumptions induced by each appliance in the household. The resulting disaggregated power time series can then be used as an input for a machine learning algorithm in order to learn consumption habits. 


Energy disaggregation (also called non-intrusive load monitoring or NILM) is a computational technique for estimating the power demand of individual appliances from a single meter which measures the combined demand of multiple appliances. The NILM problem can be formulated as follows: Let $\Bar{y}(t)$ the aggregated energy consumption measured at time $t$. With no loss of generality, $\Bar{y}(t)$ can represent the active power (The power which is actually consumed or utilised in an AC Circuit in kW). Then $\Bar{y}(t)$ can be expressed as in:

\begin{equation}
    \Bar{y}(t)=\sum_{i=1}^Cy^{(i)}(t)+e(t)
\end{equation}

Where $C$ is the number of appliances, $y^{(i)}$ the consumption induced by appliance $i$ and $e(t)$ some noise. The aim is to find $y^{(i)}$ given $\Bar{y}(t)$. 


There exist two approaches for NILM, namely load identification and source separation. In the first case, a first step called signature detection corresponds to the activation of a given appliance then a classification algorithm classifies the appliance category. The idea behind load identification is to build appropriate features called load signatures that allow to easily distinguish the referenced appliance from others within the installation. In the latter case, separation is directly obtained while retrieving the source signal. 

In order to manage a building efficiently, for example using a reinforcement learning agent, it is necessary to use a model that performs source separation while being causal. In signal processing, a causal model is a model that performs the needed task (here source separation) without looking beyond time $t$ rather than having to look in the future as presented in figure \ref{fig:dilated}. The model can then be used as a backend for prediction, which is necessary for energy management. 

\begin{figure}[h!]
    \centering
    \includegraphics[width=\columnwidth]{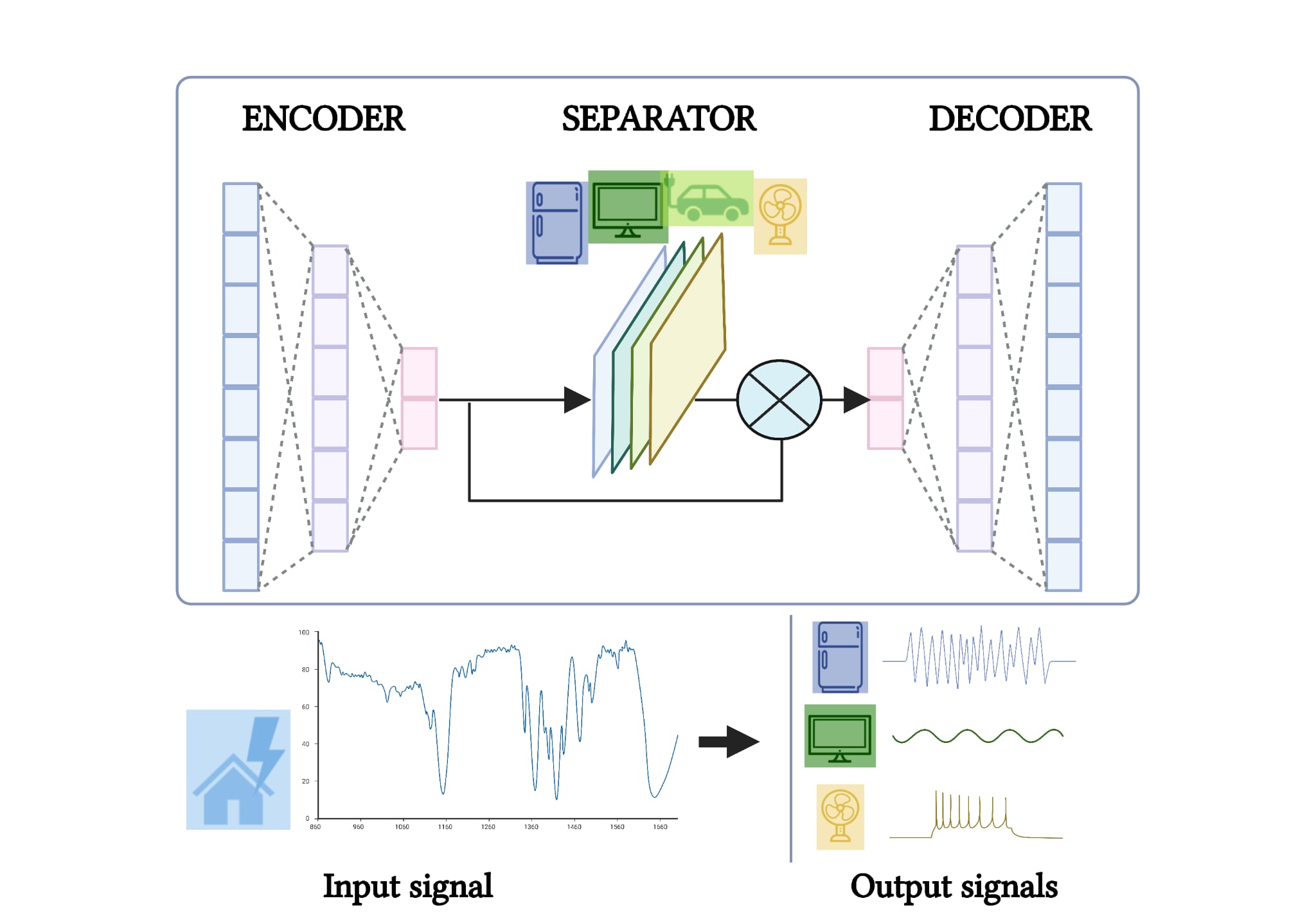}
	\caption{Overview of Conv-NILM-net. It is composed of 2 blocs, the encoder/decoder and the separator. The encoder first projects the signal into a latent space, the separator disaggregates it into $C$ corresponding to each appliance by learning  $C$ masks applied to input signal, then the decoder projects the $C$ signals to the input space}
    \label{fig:framework}
\end{figure}

We propose Conv-NILM-net, a fully convolutional and causal neural network for end-to-end energy source separation. Conv-NILM-net is inspired from Conv-TasNet \cite{tasnet}, a convolutional model for speech separation. The model does not require more quantities than active power and disaggregates the signal for multiple appliances at once. We evaluate it on REDD and UK-DALE datasets, compared to recent models and achieves state of the art performance. Figure \ref{fig:framework} presents an overview of the model and table \ref{tab:notation} summarises the notations used throughout this paper.

\begin{table*}[!ht]
	\centering
	\caption{Summary of notation.}
    \label{tab:notation}
	\begin{tabular}{ll}
		 \hline
$\bar{y}(t)$ & Aggregated energy consumption at time $t$, $\bar{y}(t) \in \mathbb R^T$\\ 
$C$ & Number of appliances\\
$T$ & Length of a signal \\
$y^{(i)}(t)$ & True energy consumption of appliance $i$ at time $t$, $y^{(i)}(t)\in \mathbb R^T$\\
$\hat{y}^{(i)}(t)$ & Predicted energy consumption of appliance $i$ at time $t$, $y^{(i)}(t)\in \mathbb R^T$\\
$Z$ & latent space representation of the mixture signal, $Z\in\mathbb R^{N\times K}$\\
 $m_i$ & learned mask for appliance $i$, $m_i\in\mathbb R^N$\\
 $s_i$ & filtered signal from the encoder for appliance i, $s_i\in\mathbb R^N$ \\
 $N$ & Number of filters in encoder-decoder\\
 $L$ & Length of the filters \\
 $B$ & Number of channels in bottleneck and the residual paths’ $1\times 1$-conv blocks\\
 $H$ & Number of channels in convolutional blocks \\
 $P$ & Kernel size in convolutional blocks \\
 $X$ & Number of convolutional blocks in each repeat \\
 $R$ & Number of repeats \\[1ex]
 \hline
\end{tabular}
\end{table*}

\section{Related Work}


Most approaches in the literature are load identification approaches that predict the state of an appliance (on/off) and predict the average consumption of the given appliance during a certain period of time. Four appliance models are usually considered:
\begin{itemize}
    \item Type I On/off devices: most appliances in households, such as bulbs and toasters;
    \item Type II Finite-State-Machines (FSM): the appliances in this category present states, typically in a periodical fashion. Examples are washer/dryers, refrigerators, and so on;
    \item Type III Continuously Varying Devices: the power of these appliances varies over time, but not in a periodic fashion. Examples are dimmers and tools.
    \item Type IV Permanent Consumer Devices: these are devices with constant power but that operate 24 h, such as alarms and external power supplies.
\end{itemize}

Current NILM methods work well for two-state appliances, but it is still difficult to identify some multi-state appliances, and even more challenging with continuous-state appliances. One of the most noticeable approaches called FHMM models each appliance as a hidden markov model (HMM) \cite{FHMM}. The HMM of each appliance is modelled independently, each one contributing to the aggregated power. AFAMAP \cite{AFAMAP} extends FHMM by predicting combinations of appliances working states. In AFAMAP, the posterior is constrained into one state change per time step. In \cite{Elastic}, the authors propose a hierarchical FHMM in order to stop imposing independence between between appliances. The algorithm takes the active power as input and performs a clustering of the correlated signals then trains an HMM on the identified clusters called super devices. During the disaggregation step, the prediction is done using AFAMAP on the super devices then the clustering is reversed to find the the original appliance. 

A critical step is the construction of load signatures or features that help to uniquely identify all types of home appliances with different operation modes. Event-based techniques have been employed to identify turn-on and turn-off events using a variety of features like the active and reactive power \cite{reactiveDL,reactive_NILM,wavenilm}, current and voltage harmonics \cite{motif,NIALM}, transient behaviour particularly during the activation and/or deactivation \cite{invest,trans}, current waveform characteristics \cite{harmonic_NN}. Although the existing harmonic-based NILM methods achieved high load identification accuracy, their applicability is limited. The main drawback of this approach is that it requires harmonic current signatures with respect to all possible combinations of devices. Consequently, the complexity of this method increases exponentially with the number of electrical devices.

Deep Learning approaches have consistently outperformed HMM-based methods. Indeed, the number of features associated with the complexity induced by all the possible devices combinations make deep learning a natural candidate for NILM. In recent years, learning-based approaches were proposed to classify and directly estimate the power consumption of type-1 and type-2 appliances from an aggregated signal. Although FHMM-based NILM approaches are extensively used for power disaggregation, their performance is limited to the accurate approximation of appliance actual power consumption especially for type-2 (multi-state) and type-4 (always-on) appliances. Moreover, HMM-based methods have been reported to suffer from scalability and generalisation, which limits its real-world application. In contrast to classical event-based and state-based approaches, deep neural networks are capable of dealing with time complexity issues, scalability issues, and can learn very complex appliance signatures if trained with sufficient data.

Most recent NILM works employing deep neural networks used 1/6-Hz or 1/3-Hz sampled active power measurement as an input feature to train various deep neural networks. such as long short-term memory (LSTM) networks \cite{LSTM,RLSTM}, denoised autoencoder \cite{neural,dae} and Convolutional Neural Networks (CNN) \cite{seqtopoint,multi_state}. \cite{neural} proposed 3 different neural networks. A convolutional layer followed by an LSTM to estimate the disaggregated signal from the global one. They also used a denoising convolutional autoencoder to produce clean signals. The last neural network estimates the beginning and the end time of each appliance activation along with the mean consumption of each. \cite{neural} performs better than FHMM however their model was unable to identify multi-state appliances. To solve the multi-state appliance identification issue, \cite{LSTM} proposed a two-layer bidirectional LSTM based DNN model. Similarly, \cite{multi_state} proposed a two-step approach to identify multi-state appliances. They used a deep CNN model to identify the type of appliances and then used a k-means clustering algorithm to calculate the number of states of appliances. 

Deep Learning also allowed source separation rather than load identification. This approach is more difficult but offers precise estimation of the consumption of each appliance in real time which includes continuous state appliances. 
\cite{seqtopoint,bidirectional} proposed sequence-to-point learning-based CNN architecture with only active power as an input feature.
In \cite{seqtoseq} gated linear unit convolutional layers \cite{GLU} are used to extract information from the sequences of aggregate electricity consumption. In \cite{DL_processing}, the authors used a deep recurrent neural network using multi-feature input space and post-processing. First, the mutual information method was used to select electrical parameters that had the most influence on the power consumption of each target appliance. Second, selected parameters were used to train the LSTM for each target appliance. Finally, a post-processing technique was used at the disaggregation stage to eliminate irrelevant predicted sequences, enhancing the classification and estimation accuracy of the algorithm.

In \cite{wavenilm}, the authors present WaveNILM which is a causal 1-D convolutional neural network inspired by WaveNet \cite{wavenet} for NILM on low-frequency data. They used various components of the complex power signal for NILM, current, active power, reactive power, and apparent power. WaveNILM, requires no significant alterations whether using
one or many input signals. However, most of the existing DNNs models for NILM use a single-task learning approach in which a neural network is trained exclusively for each appliance. That is also the case for WaveNILM. This strategy is computationally expensive and ignores the fact that multiple appliances can be active simultaneously and dependencies between them. In \cite{unet_nilm} the authors introduce UNet-NILM for multi-task appliances’ state detection and power estimation, applying a multi-label learning strategy and multi-target quantile regression. The UNet-NILM is a one-dimensional CNN based on the U-Net architecture initially
proposed for image segmentation \cite{unet}. However, this model is not causal like WaveNILM. 

Conv-NILM-net achieves the best of both worlds as it can handle source separation for any type of appliance, for multiple appliances simultaneously, it only needs the active power (although it is possible to other types of current in the same time). 

\section{Conv-NILM-Net}

The model aims at separating $C$ individual power sources $y^{(i)} \in \mathbb R^T$, where $i\in \{1,2,\dots,C\}$ from a mixture of signals representing the total consumption $\Bar{y}(t)= \sum_{i=1}^C y^{(i)}(t)+e(t)$ and $T$ is the length of the waveform. Therefore it take as input a single channel time series corresponding to the total consumption and outputs $C$ time series corresponding to the consumption of each individual appliance. In this section, we present and detail our proposed architecture. We will describe the overall structure before focusing on the separation module. 

\subsection{Overall structure}
Conv-NILM-Net is an adaptation of ConvTas-net \cite{tasnet}. Conv-Tasnet was originally only designed for speech separation and limited to two speakers. We propose an adaptation to energy load source separation with theoretically no limitation to the number of appliances. Our fully convolutional model is trainable end-to end and uses the aggregated active power as only input making the training easily deployable (no additional costly features needed).

Conv-NILM-net architecture consists of two parts: an encoder/decoder, and a separator. The encoder generates a multidimensional representation of the mixture signal; the separator learns masks applied to this representation to decompose the mixture signal, then the decoder translates the obtained signals from the encoded representation to the classic active power. The masks are found using a temporal convolutional network (TCN) consisting of stacked 1-D dilated convolutional blocks, which allows the network to model the long-term dependencies of the signal while maintaining a small model size.

Using encoder filters of legnth $L$, the model first segments the input total consumption into $K$ overlapping frames $\Bar{y}_k \in \mathbb R^L$, $k=1,2,\dots,K$ each of length $L$ with stride $S$. $\Bar{y}_k$ is transformed into a $N$-dimensional representation, $Z\in \mathbb R^{N\times K}$:
\begin{equation}
    Z = \mathcal{F}(w\cdot\Bar{Y})
\end{equation}
where $Y\in\mathbb R^{L\times K}$ and $w\in \mathbb R^{N\times K}$ the $N$ learnable basis filters of length $L$ each. $Z$ represents the latent space representation of the mixture series while $\mathcal{F}$ is a non-linear function. To ensure that the representation is non-negative, Conv-tasnet \cite{tasnet} uses the rectified linear unit (ReLU). However, this choice leads to a vanishing gradient behaviour, driving the norm of the gradients towards 0 thus making the model collapse as it eventually outputs null signals. Therefore, we replace ReLU with Leaky ReLU and only use ReLU for the last layer of the separation masks to enforce positive outputs. 

The separator predicts a representation for each source by learning a mask in this latent space. It is performed by estimating $C$ masks $m_i\in \mathbb R^N$. The representation of each source $s_i \in \mathbb R^N$, is then calculated by applying the corresponding mask $m_i$ to the mixture representation, using element-wise multiplication:
\begin{equation}
s_i=Z \odot m_i
\end{equation}

In Conv-tasnet as well as in \cite{tas}, the masks were constrained such that $\sum_{i=1}^Cm_i=1$. This was applied based on the assumption that the encoder-decoder  architecture can perfectly reconstruct the input mixture. indeed, in their model, $e(t)=0$, $\forall t$. This assumption cannot be made in a NILM context, it is therefore relaxed.  

The input signal of each source is then reconstructed by the decoder:
\begin{equation}
    \hat{y}^{(i)} = s_i\cdot V
\end{equation}

where $V \in \mathbb R^{N\times L}$ are the decoder filters, each of length $L$. 

\subsection{Separation module}

\begin{figure}
    \centering
    \begin{minipage}{0.45\textwidth}
        \centering
        \includegraphics[width=0.9\textwidth]{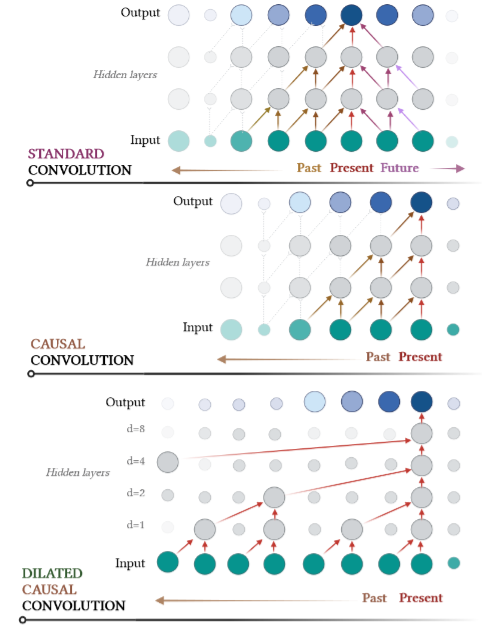} 
        \caption{Causal and dilated convolutions.}
        \label{fig:dilated}
    \end{minipage}\hfill
    \begin{minipage}{0.45\textwidth}
        \centering
        \includegraphics[width=0.9\textwidth]{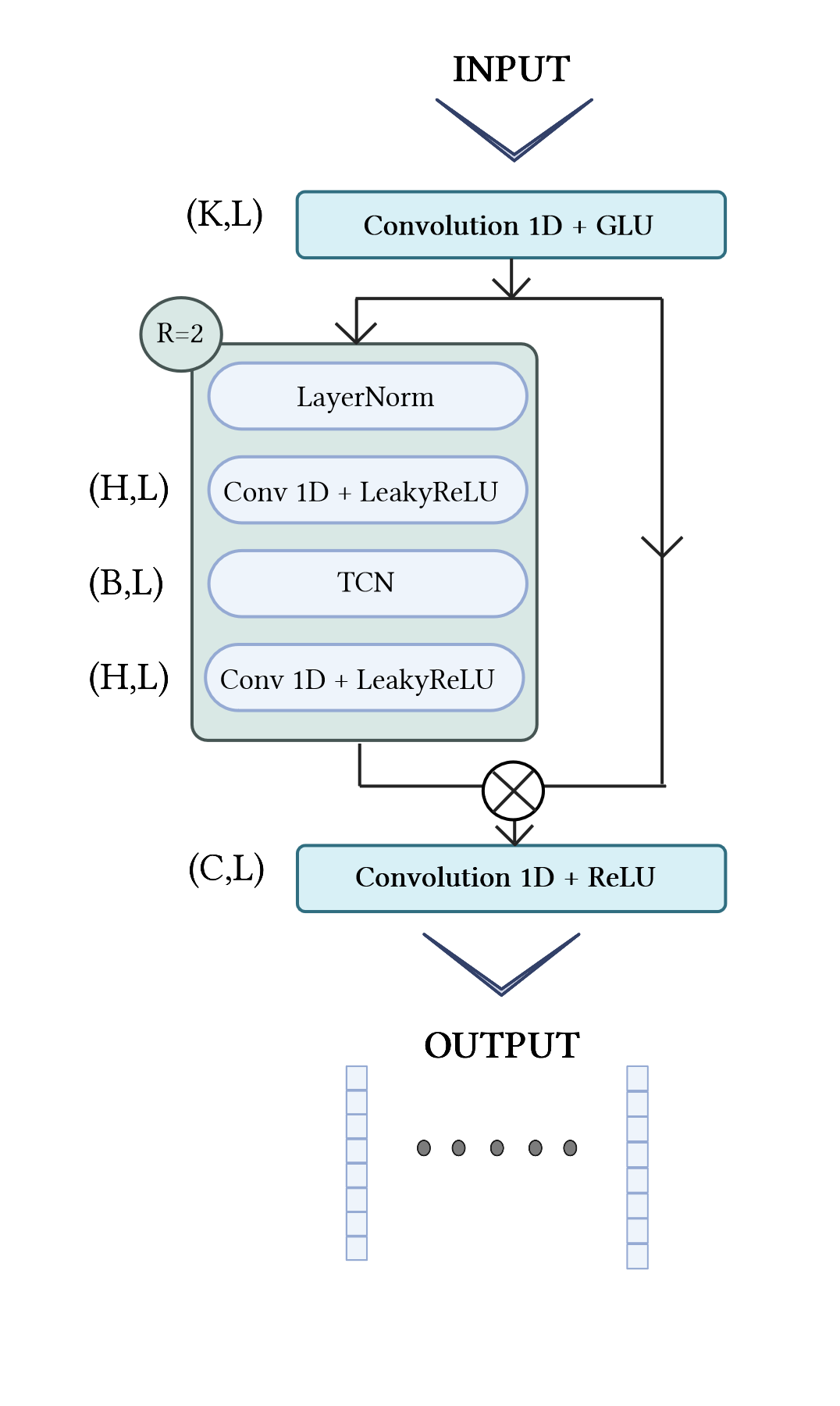} 
        \caption{Detailed representation of the separator.}
        \label{fig:fig_NILM}
    \end{minipage}
\end{figure}

The separator is mainly based on a temporal convolutional network (TCN) \cite{TCN} and is detailed in figure \ref{fig:fig_NILM}. Temporal convolutions require the use of dilated convolutions which aim to increase the receptive field. Indeed, pooling or strided convolutions are usually implemented for this purpose, however they reduce the resolution. Dilated convolutions allow exponential expansions of the receptive field without loss of resolution, while achieving same computation and memory costs. These are simply implemented by defining a spacing between the values in a kernel as illustrated in figure \ref{fig:dilated}. 

In \cite{LSTM}, the authors used LSTM \cite{Lsurvey} for NILM. This architecture can handle long sequences but suffers from the vanishing gradient issue while being computationally costly. We argue that a more efficient approach is to make use of 1D-convolutions. As illustrated in figure \ref{fig:dilated}, convolutions for time series require future values (compared to the point of reference). During inference, these values are not accessible making the use of this model unpractical for prediction. This can be solved by giving a causal formulation to convolutions where the present value only depends of past ones. Moreover, the implementation is easy as it only requires an asymetric padding. 

However relevant values become sparse. As in \cite{tasnet}, Conv-NILM-net uses dilated layers with exponentially increasing dilation rates. The dilation factors increase exponentially to ensure a sufficiently large temporal context window to take advantage of the long-range dependencies in the signal. Therefore the dilation factors increase exponentially to ensure a sufficiently large temporal context window to take advantage of the long-range dependencies in the signal. Therefore TCN consists of 1-D convolutional blocks with increasing dilation factors. Given kernels of length $L$ and $l$ layers, the receptive field of Conv-NILM-net is of size $RF=2^l(L-1)$. 

The output of the TCN is passed to a $1\times1$ conv block for mask estimation. This block also serves for dimensionality reduction and together with a nonlinear activation function estimates $C$ mask vectors for the $C$ target sources. The last layer of the last bloc uses a ReLU activation function to ensure the non-negativity of the outputs. 

Contrary to speech separation, where simultaneous speeches are independent from one another, it is not the case in NILM context where appliance activations can be  highly dependent. An elegant solution proposed in \cite{emp,seqtoseq}, can be to use gated linear units (GLU) \cite{GLU} to replace LeakyReLU activation functions. GLU allow the model to decide itself the relative importance of the kernels by using two parallel convolutions with the first followed by a sigmoid which result is then multiplied with the second convolution. The output of the sigmoid acts like a mask that activates depending on the input of the second convolution. 

\begin{figure}[h!]
    \centering
    \includegraphics[width=0.7\columnwidth]{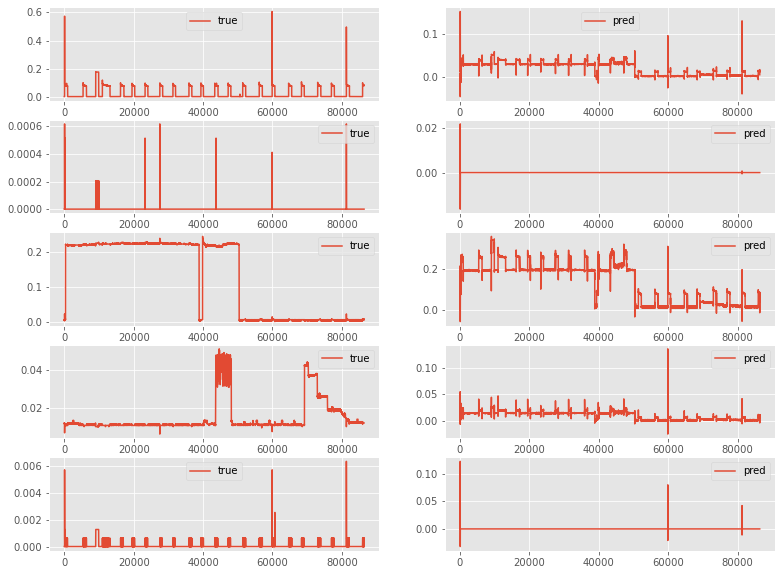}
	\caption{Outputs of Conv-NILM-net of top 5 appliances of REDD building 1 when trained using classic MSE. We observe the presence of the artefacts although the MSE is minimized.}
    \label{fig:MSE}
\end{figure}

MSE, $L_1$, or even SI-SNR \cite{tasnet} losses are often used for NILM or source separation problems. The MSE takes the average squared error on all time steps for all disaggregated signals (ie appliances). We found that taking the mean on appliances is detrimental to the learning process as the error is distributed over all appliances. Therefore, the signals get mixed and artefacts of most consuming appliances appear on the remaining ones as illustrated in figure \ref{fig:MSE}. We therefore choose to sum the error over all appliance rather than averaging it. The window mean squared error is calculated as:

\begin{equation}
    \text{WMSE}=\frac{1}{T}\sum_{t=0}^T\sum_{i=1}^C\left(\hat{y}^{(i)}(t)-\bar{y}^{(i)}(t)\right)^2
\end{equation}

\section{Experimental Methodology, Results, and Discussion}

\subsection{Datasets and parameters}
Our experiments are done on two real-world datasets, Reference Energy Disaggregation Dataset (REDD) \cite{redd} and UK-DALE \cite{UK-DALE}.  REDD records the power for 6 houses with sampling frequency of 1Hz for mains meter and 1/3Hz for appliance-channel meters. We choose to disaggregate the five top appliances for each building. UK-DALE data set published in 2015, records the power demand from five UK houses. In each house we record both the whole-house mains power demand every six seconds as well as power demand from individual appliances every six seconds. 

For REDD, we converted the disaggregated data to 1Hz using linear extrapolation and kept $1/6$ Hz frequency for UK-DALE. Usually the data are normalised and the series of each appliance are scaled individually. We argue that appliance scaling is not practical as it is not possible to apply the scaling factors to the global signal available during inference. Therefore we chose to only use a minmax scaler for all appliances combined directly on the mixture power signal. In some contributions like seq2seq and seq2point \cite{bidirectional,seqtopoint}, a sliding window of the aggregate power is used as the input sequence, and the midpoint of the window corresponding to the target device reading is used as the output. This preprocessing smooths the power loads and makes the target values to retrieve easier. All results presented in section \ref{results} for these implementations were obtained using smoothing. The results presented for Conv-NILM-NET were obtained without smoothing, making the difference in performance even more noticeable. 

For UK-DALE dataset, we compare our results to UNET-NILM \cite{unet_nilm} and seq2point \cite{bidirectional}. The constructed artificial aggregate consumption is obtained by taking the summation of selected appliances plus additional one appliance (Television in this setting). For UNET-NILM, the authors used a quantile filter to smooth the signal. This is not required for Conv-NILM-Net. 

In our implementation for Conv-NILM-Net, we used for, each dataset, one day as input. This means that with 1Hz frequency, the input to Conv-NILM-net was 86400 points for REDD and 14400 for UK-DALE ($1/6$ Hz). 
The used parameters for Conv-NILM net are: $N=32$; $L=48$; $B=2$; $H=P=X=3$; $R=2$. The meaning of each notation is made available in table \ref{tab:notation} where we kept the same notation as in \cite{tasnet}. The model was trained for 2000 epochs using 10-fold cross-validation and a batch size of $5$. We used Adam optimiser with an initial learning rate $lr=0.01$, $\text{betas}=(0.9, 0.999)$, $\text{eps}=0.01$. 
 
\subsection{Metrics}

We evaluate the performance of the framework using the mean absolute error (MAE). Estimated accuracy is also a common metric for evaluating disaggregated power.

\begin{equation}
    \text{Est.Acc.}=1-\frac{\sum_{t=1}^T\sum_{i=1}^C|\hat{y}^{(i)}(t)-\bar{y}^{(i)}(t)|}{2\sum_{t=1}^T\sum_{i=1}^C\bar{y}^{(i)}(t)}
\end{equation}

Where $\hat{y}^{(i)}(t)$ is the predicted power level of appliance $i$ at time $t$, and $\bar{y}^{(i)}(t)$ is the ground truth. The above expression yields total estimated accuracy; if needed, the summation over $i$ can be removed creating an appliance-specific estimation accuracy. We also report the Signal Aggregate Error (SAE):

\begin{equation}
    \text{SAE} = \frac{|\hat{r}-r|}{r}
\end{equation}

where $\hat{r}$ and r represent the predicted total energy consumption of an appliance and the ground truth one. SAE measures the total error in the energy within a period, which is accurate for reporting daily power consumption even if its consumption is inaccurate in every time point.

\subsection{Results on REDD}\label{results}

Table \ref{tab:conv-redd} presents the results obtained on REDD dataset for five building. For each building with disaggregated the top five appliances and reported the MAE, estimated accuracy ans SAE. We tested 3 versions on Conv-NILM-net. We observe that the causal+GLU tend to perform better on average but its results are very close to the causal implementation while increasing the number of parameters dramatically. We therefore tend to prefer the causal version of our model. 

\begin{table}
\centering
\begin{tabular}{CCCCCCCCCCC}
\toprule
\multicolumn{1}{l}{ } &
\multicolumn{1}{l}{ } &
\multicolumn{9}{c}{Model}   \\ 
\cmidrule(lr){3-11}

&
&
\multicolumn{3}{c}{Conv-NILM-NET} &
\multicolumn{3}{c}{Causal}     &
\multicolumn{3}{c}{Causal + GLU} \\
\cmidrule(lr){3-5}
\cmidrule(lr){6-8}
\cmidrule(lr){9-11}

\multicolumn{1}{c}{Building} &
\multicolumn{1}{c}{Appliance} &
\multicolumn{1}{c}{MAE} &
\multicolumn{1}{c}{Est.Acc} &
\multicolumn{1}{c}{SAE} &
\multicolumn{1}{c}{MAE} &
\multicolumn{1}{c}{Est.Acc} &
\multicolumn{1}{c}{SAE} &
\multicolumn{1}{c}{MAE} &
\multicolumn{1}{c}{Est.Acc} &
\multicolumn{1}{c}{SAE} \\
\midrule

\textbf{Building 1} & 
    \begin{tabular}{ c } Fridge  \\ 
     Washer dryer\\ Light \\ Sockets \\ Dishwasher \\ \textbf{Total}\\  \end{tabular} &
    \begin{tabular}{ c } 0.049 \\
     0.005 \\ 0.063\\ 0.015\\ 0.006\\ 0.027\\
     \end{tabular} & 
     \begin{tabular}{ c } 0.900 \\
     0.937 \\ 0.970\\ 0.874\\ 0.916\\ 0.919\\
     \end{tabular} & 
     \begin{tabular}{ c } 0.058 \\
     0.074 \\ 0.030\\ 0.092\\ 0.102\\ 0.071\\
     \end{tabular} &
     \begin{tabular}{ c } 0.006 \\
     0.002 \\ 0.007\\ 0.005\\ 0.002\\0.005\\
     \end{tabular} & 
     \begin{tabular}{ c } 0.981 \\
     0.993 \\ 0.980\\ 0.984\\ 0.993\\0.986\\
     \end{tabular} & 
     \begin{tabular}{ c } 0.053 \\
     0.088 \\ 0.021\\ 0.131\\ 0.095\\0.078\\
     \end{tabular} &
     \begin{tabular}{ c } 0.004 \\
     0.002 \\ 0.008\\ 0.006\\ 0.003\\\textbf{0.004}\\
     \end{tabular} & 
     \begin{tabular}{ c } 0.987 \\
     0.990 \\ 0.984\\ 0.992\\ 0.997\\\textbf{0.989}\\
     \end{tabular} & 
     \begin{tabular}{ c } 0.051 \\
     0.103 \\ 0.033\\ 0.191\\ 0.070\\\textbf{0.054}\\
     \end{tabular}\\
     \hline
\textbf{Building 2} & 
    \begin{tabular}{ c } Fridge  \\ 
     Washer dryer\\ Light \\ Sockets \\ Dishwasher \\ \textbf{Total}\\  \end{tabular} &
    \begin{tabular}{ c } 0.038 \\
     0.018 \\ 0.012\\ 0.002\\ 0.0006\\\textbf{0.014}\\
     \end{tabular} & 
     \begin{tabular}{ c } 0.912 \\
     0.914 \\ 0.986\\ 0.993\\ 0.997\\ 0.960\\
     \end{tabular} & 
     \begin{tabular}{ c } 0.052 \\
     0.109 \\ 0.024\\ 0.059\\ 0.115\\0.0718\\
     \end{tabular} &
     \begin{tabular}{ c } 0.032 \\
     0.031 \\ 0.002\\ 0.001\\ 0.001\\\textbf{0.014}\\
     \end{tabular} & 
     \begin{tabular}{ c }0.939 \\
     0.940 \\ 0.981\\ 0.991\\ 0.993\\ 0.969\\
     \end{tabular} & 
     \begin{tabular}{ c } 0.068 \\
     0.098 \\ 0.104\\ 0.032\\ 0.029\\\textbf{0.066} \\
     \end{tabular} &
    \begin{tabular}{ c } 0.041 \\
     0.034 \\ 0.008\\ 0.006\\ 0.003\\0.018 \\
     \end{tabular} & 
     \begin{tabular}{ c } 0.956 \\
     0.967 \\ 0.987\\ 0.990\\ 0.992\\\textbf{0.978} \\
     \end{tabular} & 
     \begin{tabular}{ c } 0.054 \\
     0.087 \\ 0.099\\ 0.058\\ 0.031\\\textbf{0.066} \\
     \end{tabular} \\
     \hline
\textbf{Building 3} & 
    \begin{tabular}{ c } Fridge  \\ 
     Washer dryer\\ Light \\ Sockets \\ Dishwasher \\ \textbf{Total}\\  \end{tabular} &
    \begin{tabular}{ c } 0.006 \\
     0.007 \\ 0.009\\ 0.009\\ 0.007\\\textbf{0.006}\\
     \end{tabular} & 
     \begin{tabular}{ c } 0.820 \\
     0.997 \\ 0.863\\ 0.961\\ 0.960\\0.920\\
     \end{tabular} & 
     \begin{tabular}{ c } 0.089 \\
     0.061 \\ 0.123\\ 0.091\\ 0.085\\ 0.90\\
     \end{tabular} &
     \begin{tabular}{ c } 0.072 \\
     0.003 \\ 0.037\\ 0.040\\ 0.032\\0.037\\
     \end{tabular} & 
     \begin{tabular}{ c } 0.822 \\
     0.993 \\ 0.854\\ 0.941\\ 0.940\\0.910\\
     \end{tabular} & 
     \begin{tabular}{ c } 0.078 \\
     0.065 \\ 0.098\\ 0.080\\ 0.096\\\textbf{0.083}\\
     \end{tabular} & 
     \begin{tabular}{ c } 0.009 \\
     0.004 \\ 0.006\\ 0.007\\ 0.005\\\textbf{0.006}\\
     \end{tabular} & 
     \begin{tabular}{ c } 0.856 \\
     0.993 \\ 0.900\\ 0.942\\ 0.989\\ \textbf{0.936}\\
     \end{tabular} & 
     \begin{tabular}{ c } 0.068 \\
     0.087 \\ 0.098\\ 0.126\\ 0.078\\0.091\\
     \end{tabular}\\
     \hline
\textbf{Building 4} & 
    \begin{tabular}{ c } Fridge  \\ 
     Washer dryer\\ Light \\ Sockets \\ Dishwasher \\ \textbf{Total}\\  \end{tabular} &
    \begin{tabular}{ c } 0.003 \\
     0.002\\ 0.002\\ 0.001\\ 0.0006\\\textbf{0.002}\\
     \end{tabular} & 
     \begin{tabular}{ c } 0.961 \\
     0.947 \\ 0.936\\ 0.981\\ 0.994\\0.964\\
     \end{tabular} & 
     \begin{tabular}{ c } 0.054 \\
     0.105 \\ 0.076\\ 0.098\\ 0.132\\0.093\\
     \end{tabular} &
     \begin{tabular}{ c } 0.007 \\
     0.016 \\ 0.019\\ 0.008\\ 0.002\\ 0.011\\
     \end{tabular} & 
     \begin{tabular}{ c } 0.982 \\
     0.933 \\ 0.940\\ 0.980\\ 0.995\\ \textbf{0.966}\\
     \end{tabular} & 
     \begin{tabular}{ c } 0.021 \\
     0.087 \\ 0.055\\ 0.016\\ 0.098\\ 0.055\\
     \end{tabular} &
     \begin{tabular}{ c } 0.050 \\
     0.031 \\ 0.002\\ 0.006\\ 0.020\\0.022\\
     \end{tabular} & 
     \begin{tabular}{ c } 0.930 \\
     0.954 \\ 0.901\\ 0.937\\ 0.892\\ 0.923\\
     \end{tabular} & 
     \begin{tabular}{ c } 0.043 \\
     0.056 \\ 0.080\\ 0.024\\ 0.069\\\textbf{0.054}\\
     \end{tabular}\\
     \hline
\textbf{Building 5} & 
    \begin{tabular}{ c } Fridge  \\ 
     Washer dryer\\ Light \\ Sockets \\ Dishwasher \\ \textbf{Total}\\  \end{tabular} &
    \begin{tabular}{ c } 0.003 \\
     0.003 \\ 0.0006\\ 0.0006\\ 0.0008\\\textbf{0.002}\\
     \end{tabular} & 
     \begin{tabular}{ c } 0.883 \\
     0.992 \\ 0.999\\ 0.966\\ 0.913\\ 0.950\\
     \end{tabular} & 
     \begin{tabular}{ c } 0.005 \\
     0.001 \\ 0.0001\\ 0.002\\ 0.004\\ \textbf{0.002}\\
     \end{tabular} &
     \begin{tabular}{ c } 0.005 \\
     0.0009 \\ 0.0001\\ 0.002\\ 0.004\\ \textbf{0.002}\\
     \end{tabular} & 
     \begin{tabular}{ c } 0.880 \\
     0.993 \\ 0.999\\ 0.97\\ 0.91\\ 0.95\\
     \end{tabular} & 
     \begin{tabular}{ c } 0.004 \\
     0.001 \\ 0.002\\ 0.005\\ 0.002\\\textbf{0.002}\\
     \end{tabular} &
     \begin{tabular}{ c } 0.005 \\
     0.003 \\ 0.002\\ 0.007\\ 0.010\\ 0.005\\
     \end{tabular} & 
     \begin{tabular}{ c } 0.991 \\
     0.983 \\ 0.977\\ 0.931\\ 0.990\\ \textbf{0.974}\\
     \end{tabular} & 
     \begin{tabular}{ c } 0.093 \\
     0.012 \\ 0.10\\ 0.037\\ 0.078\\ 0.064\\
     \end{tabular}\\
\bottomrule
\end{tabular}
\caption{Conv-NILM-Net scaled results on top five appliances REDD dataset. Best average results are highlighted in bold.}
\label{tab:conv-redd}
\end{table}

Table \ref{REDD} compares the performance of Conv-NILM-net with state of the art models on 3 appliances that appear on REDD dataset. These appliances were selected as they are the only one presented in \cite{CNN-DI}. We therefore were limited to these appliance to compare our framework. We observe that our models outperform the state of the art by a margin. It decreases the MAE by $45\%$ for the fridge, $51\%$ for the microwave and even by $80\%$ for the dishwaser. The best performing model is the causal model. In the appendix we present some outputs of the model for buildings 1 to 4 from REDD. These were obtained when disaggregating the top 5 appliances detailed in the same order as in table \ref{tab:conv-redd}.

\begin{table}
\centering
\begin{tabular}{CCCC}
\hline
\textbf{Model} & \textbf{Fridge} & \textbf{Microwave} & \textbf{Dishwasher}\\
\hline
    \begin{tabular}{ c } seq2point  \\ 
     seq2seq\\ GLU-Res \cite{seqtoseq} \\ CNN-DI\cite{CNN-DI} \\ Conv-NILM-NET \\ Conv-NILM-NET (causal) \\ Conv-NILM-NET (GLU, causal)\\ \end{tabular} &
    \begin{tabular}{ c } 28.104 \\
     30.63 \\ 21.97\\ 26.801\\ 14.67\\14.21\\15.02\\
     \end{tabular} & 
     \begin{tabular}{ c } 28.199 \\
     33.272 \\ 25.202\\ 19.455\\ 9.67\\8.51\\9.76\\
     \end{tabular} & 
     \begin{tabular}{ c } 20.048 \\
     19.449 \\ 33.37\\ 17.665\\ 3.56\\3.29\\3.31\\
     \end{tabular}\\
     \hline
\end{tabular}
\caption{MAE results for Buiding 1 of REDD dataset. }
\label{REDD}
\end{table}

\subsection{Results on UK-DALE}

Table \ref{tab:UK-DALE} compares the MAE of our model on UK-DALE dataset with UNET-NILM and seq2point. Our model outperforms the state of the art on the selected appliances. The causal model performs the best again while the total average is decreased by $33\%$ compared to seq2pont. 

\begin{table}
    \centering
\begin{tabular}{CCCCCCC}
\toprule
\multicolumn{1}{l}{Appliance} &
\multicolumn{6}{c}{Model}   \\ 
\cmidrule(lr){2-7}

&
\multicolumn{1}{c}{1D-CNN} &
\multicolumn{1}{c}{UNET-NILM}     &
\multicolumn{1}{c}{Seq2point} & 
\multicolumn{1}{c}{Ours} & 
\multicolumn{1}{c}{Ours (causal)} &
\multicolumn{1}{c}{Ours (causal + GLU)} \\
\midrule
\text{Kettle} & 20.390 & 16.003 & 2.16 & 1.85 & 1.9 & 2.5\\
\text{Freezer} & 18.583 & 15.124 & 8.136 & 5.32 & 5.01 & 6.1\\
\text{Dish washer} & 9.884 & 6.764 & 3.49 & 2.42 & 2.01 & 2.55\\
\text{Washing machine} & 15.758 & 11.506 & 4.063 & 2.3 & 2.15 & 2.39\\
\text{Microwave} & 9.690 & 6.475 & 1.305 & 0.902 & 0.91 & 1.05\\
\text{\textbf{Total}} & 14.86 & 11.174 & 3.831 & 2.56 & 2.40 & 2.918\\
\bottomrule
\end{tabular}
\caption{Experimental results (MAE) in the UK-DALE dataset.}
\label{tab:UK-DALE}
\end{table}

Table \ref{tab:params} compares the size of Conv-NILM-net with state of the art models in terms of number of parameters. We observe that the fully convolution architecture of our model along with its particular architecture (encoder/decoder + separator) allow to obtain state of the art results with a model of approximately 40K parameters. This also possible because, contrary to other models, we use a unique loss for only one task. For instance UNET-NILM uses two separate loss functions, one to detect activation and an other to regress the average consumption while Seq2point \cite{bidirectional} uses bidirectional residual networks which are very deep. It is also valuable to notice that models like UNET-NILM are specialized on individual appliances, meaning that in one needs to disaggregate 5 appliances, it requires 5 models, multiplying the number of parameters.  

\begin{table}
\centering
\begin{tabular}{CC}
\text{Models} & \text{\# parameters} \\
\hline
\text{seq2point} & 29.2M \\
\text{seq2seq} & 29.8M \\
\text{GLU-Res} & 1.2M \\
\text{CNN-DI} & 738K \\
\text{Conv-NILM-net} & 41088 \\
\hline
\end{tabular}
\caption{number of parameters}
\label{tab:params}
\end{table}

\begin{figure*}[h!]
        \centering
        \begin{subfigure}[b]{0.475\textwidth}
            \centering
            \includegraphics[width=\textwidth]{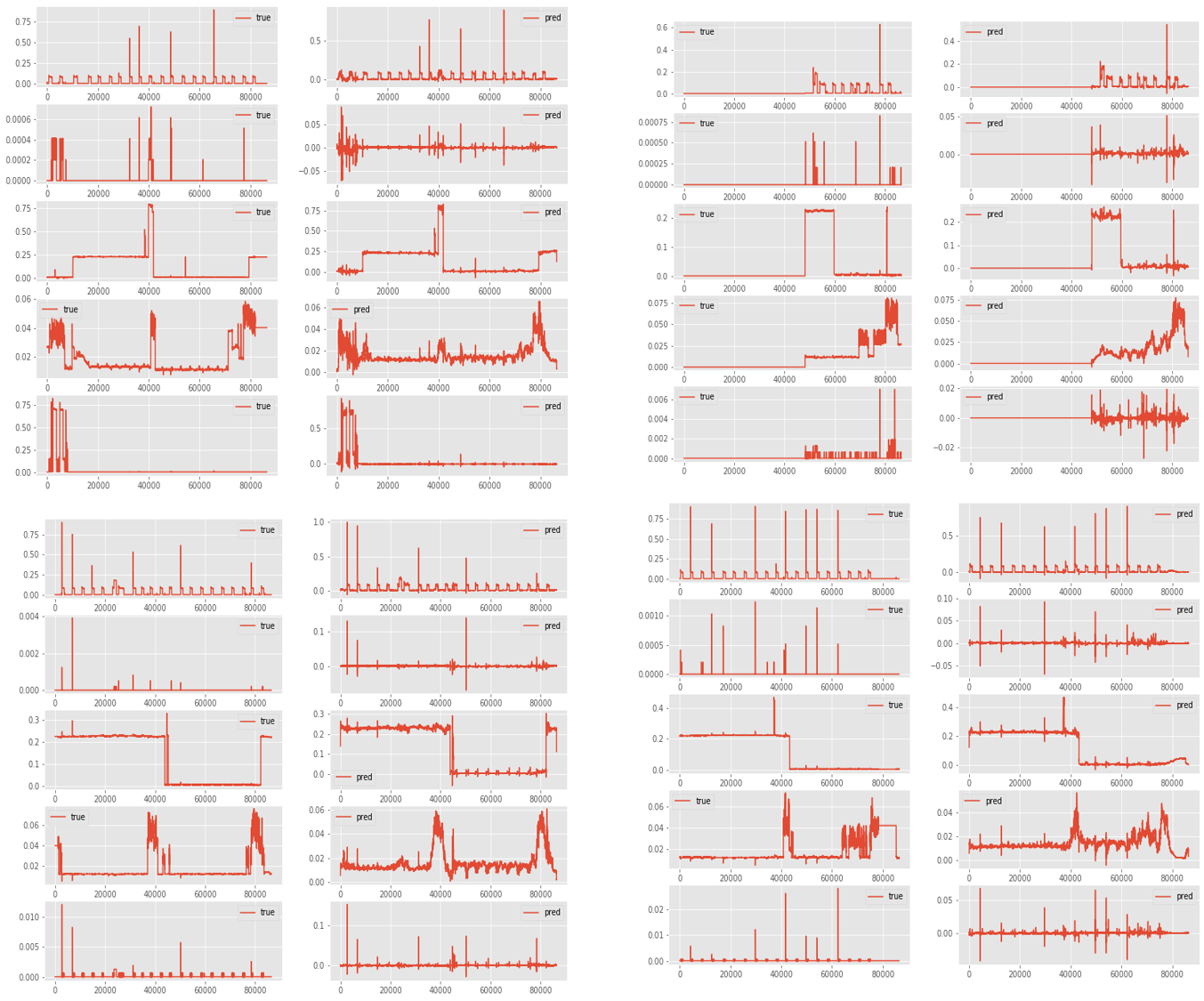}
            \subcaption{building 1}
        \end{subfigure}
        \hfill
        \begin{subfigure}[b]{0.475\textwidth}  
            \centering 
            \includegraphics[width=\textwidth]{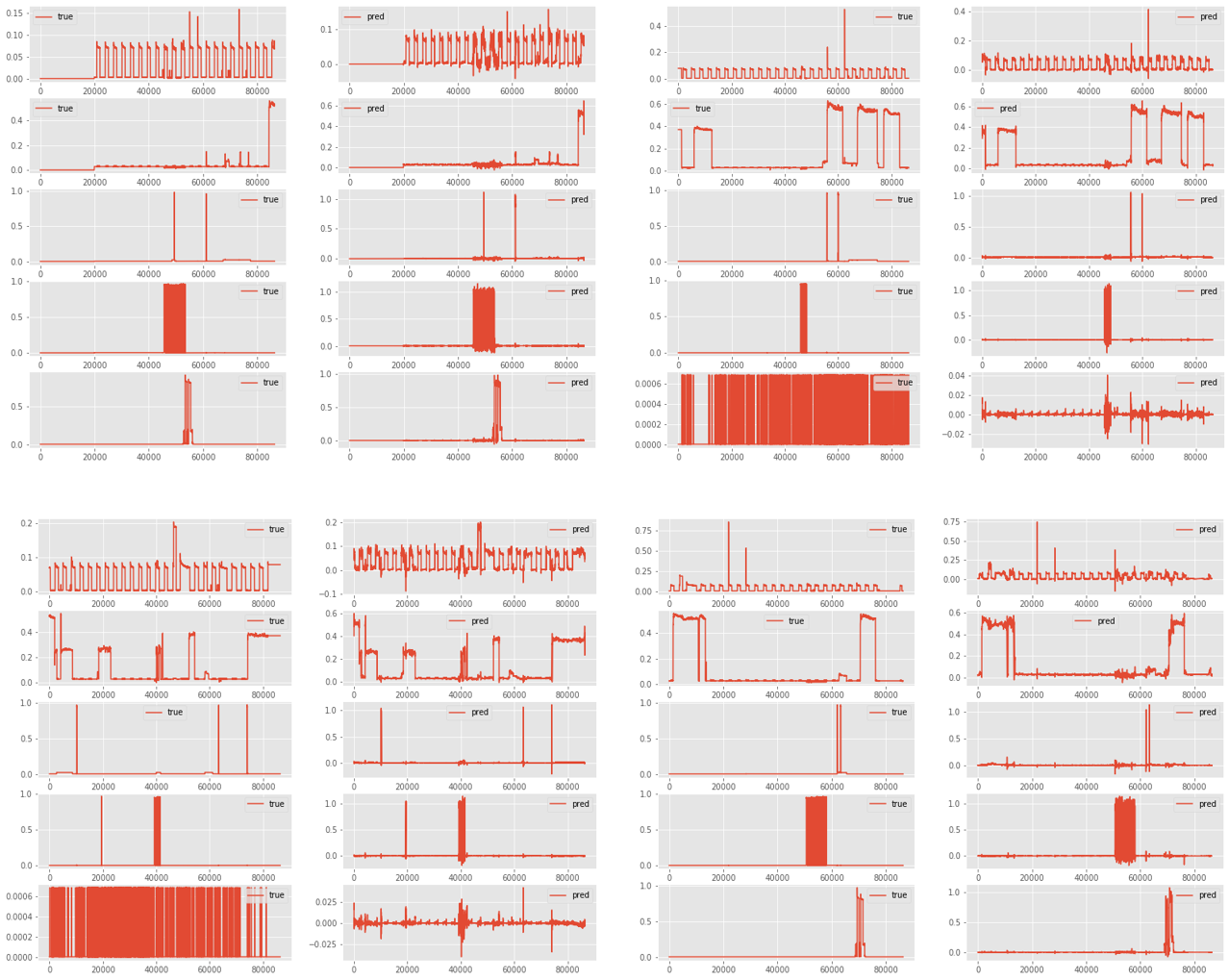}
            \subcaption{Building 2}
        \end{subfigure}
        \vskip\baselineskip
        \begin{subfigure}[b]{0.475\textwidth}   
            \centering 
            \includegraphics[width=\textwidth]{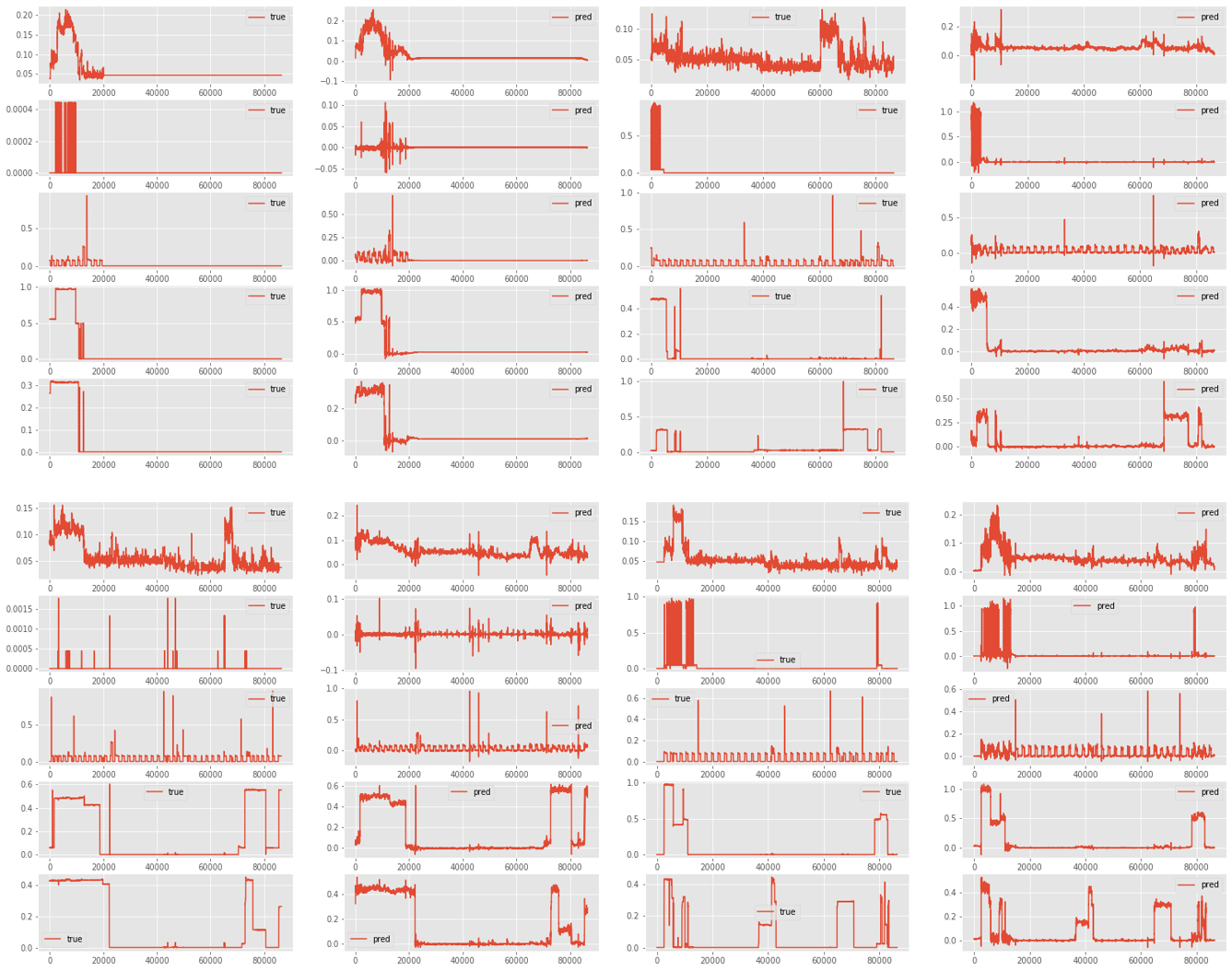}
            \subcaption{Building 3}
        \end{subfigure}
        \hfill
        \begin{subfigure}[b]{0.475\textwidth}   
            \centering 
            \includegraphics[width=\textwidth]{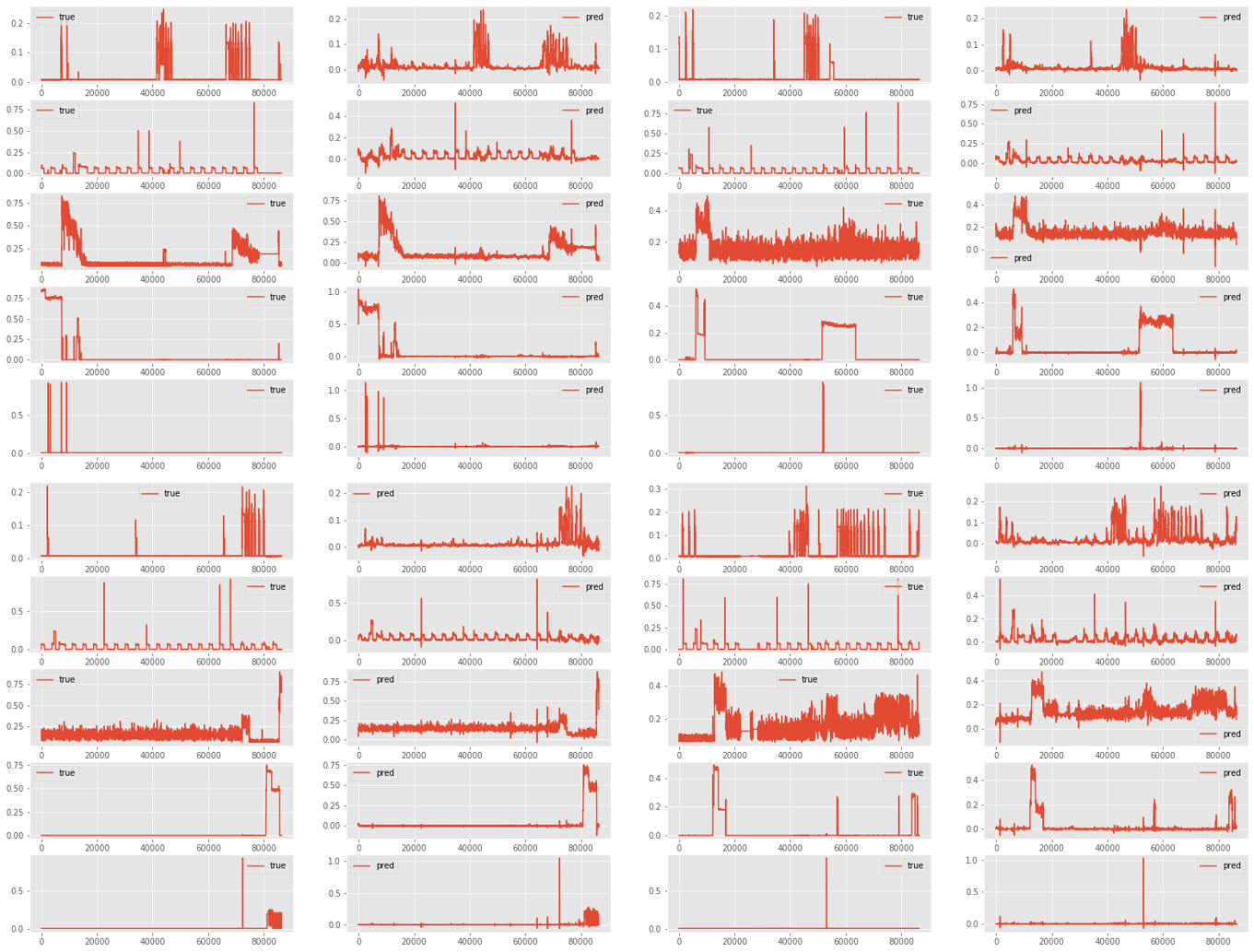}
            \subcaption{Building 4}
        \end{subfigure}
        \caption{Selected results on top five appliances of first 4 building of REDD dataset. For each building the appliances are presented in the same order as in table \ref{tab:conv-redd}. The left panels correspond to the disaggregated target signals and the right panels presents the predicted output from Conv-NILM-net.} 
        \label{fig:some_results}
    \end{figure*}

Finally, figure \ref{fig:some_results} presents some results on top five appliances of first 4 building of REDD dataset. For each building the appliances are presented in the same order as in table \ref{tab:conv-redd}. The left panels corresponds to the disaggregated target signals and the right panels presents the predicted output from Conv-NILM-net. 

\section{Conclusion}

In this work, we presented Conv-NILM-net, an adaptation of Convtas-net to non intrusive load monitoring. We tested our model on two real world dataset and showed that Conv-NILM-net outperforms the state of the art by a margin. We presented 2 alternate models, one being causal and other using Gated Linear Units (GLU). These models allowed accurate disaggregation of several appliances at once while being much more smaller than their existing counterparts. Finally, the causal model allows
consumption prediction and is ideal as input to an energy management system or a reinforcement learning model. In future work, we will use causal conv-NILM-net as a prediction model and test it in a reinforcement learning context. We will also test the GLU augmented model to verify if this implementation effectively takes into account appliances inter-dependencies and helps achieve better consumption predictions.

\bibliographystyle{plain}
\bibliography{Bibliography.bib}

\end{document}